\documentclass[12pt]{article}
\usepackage[dvips]{graphicx}
\usepackage{amsmath}
\usepackage{amsfonts}
\usepackage{latexsym}
\usepackage{amssymb}
\usepackage{multicol}
\usepackage{array}
\newcommand{\be}{\begin{equation}}
\newcommand{\ee}{\end{equation}}
\newcommand{\bea}{\begin{eqnarray*}}

\newcommand{\eea}{\end{eqnarray*}}
\newcommand{\pa}{\partial}
\renewcommand{\ll}{\left(}
\newcommand{\rr}{\right)}

\newcommand{\sy}{supersymmetry}
\newcommand{\sg}{supergravity}
\newcommand{\dr}{D\!\!\!\!/}
\def\a{\alpha}
\def\b{\beta}
\def\c{\chi}
\def\d{\delta}
\def\e{\epsilon}
\def\f{\phi}
\def\g{\gamma}

\def\j{\psi}

\def\l{\lambda}
\def\m{\mu}
\def\n{\nu}
\def\o{\omega}

\def\q{\theta}
\def\r{\rho}
\def\s{\sigma}

\def\v{\varphi}
\def\x{\xi}

\begin{document}

\centerline{\LARGE\bf Supergravity as a Yang-Mills theory\footnote{Contribution to ``50 Years of Yang-Mills Theory'', World Pub. Co., G. `t Hooft editor.}}

\vspace{1cm}
\centerline{
\bf{Peter van Nieuwenhuizen}
}
\vspace{.5cm}
\centerline{
{\it C.N. Yang Institute for Theoretical Physics,} }
\centerline{\it State University of New York at Stony Brook,
NY 11794-3840, USA}

\begin{center}
{\bf Abstract}: We give a simple introduction to ordinary and conformal 
supergravity, and write their actions as squares of curvatures. 
\end{center}

\section{Introduction}
Supergravity is the nonabelian gauge theory of supersymmetry. It was
constructed in 1976~\cite{1,2}, soon after rigid \sy\  had been
constructed in the early 1970's~\cite{3}. There already exist many books
and reviews on this subject, so in this contribution we
shall not try to give a systematic account of supergravity, but
rather put the work of Yang and Mills central, and focus on the
similarities and differences between \sg\ and Yang-Mills theory.
Let us only mention a few of the successes of supergravity.\\
$\bullet$ It is a complete classical theory in the same way as general 
relativity, with action and well-understood
geometry,
and forms the low-energy limit of string theory where such complete
results have not yet been found\\
$\bullet$ it allows a proof of the positive energy conjecture in general
relativity\\
$\bullet$ it has given relations between 5-dimensional
tree supergravity and 4-dimensional nonperturbative quantum
superYang-Mills theory (the AdS/CFT correspondence)\\
$\bullet$ it has made the phenomenology of the search for supersymmetric
particles
at LHC possible, because it can remove the huge cosmological constant
of spontaneously broken rigid supersymmetry and the gravitino can eat
a Goldstino, explaining why no Goldstino has been seen \\
$\bullet$ it has led to various dualities in field theory which
one later attempted to extend to string theory\\
$\bullet$ the unique 11-dimensional supergravity theory provides the 
only
concrete information about the hypothetical  M-theory which is supposed
to describe all non-perturbative string theory including solitons\\
$\bullet$  it rephrases differential geometry in terms of Killing spinors
instead of Killing vectors. This approach is 
much more powerful and has led to breakthroughs
in various areas of mathematics.

Yang-Mills theory, the gauge theory of internal nonabelian symmetries,
has
become in the 20-th century what Maxwell theory was in the 19-th
century. Its renormalization by 't Hooft and Veltman has led to a
consistent quantum gauge field theory, whose radiative corrections
are computed and measured in the large laboratories all over the
world. The results confirm the theory to incredible precision.  As
the underlying theory for QCD and electroweak forces, it has
resolved the problems of earlier approaches such as the $V-A$
theory of the weak interactions, or the one-boson exchange models,
bootstrap models, dispersion relation approaches, Regge models,
etc. for the strong interactions. A theoretical foundation for
the Standard Model has been established through the work of Yang,
Mills, 't Hooft, Veltman, and  Faddeev, Popov,
Fradkin, Tyurin, Feynman, Gell-Mann, Bryce DeWitt, Mandelstam,
Slavnov, Taylor, Zinn-Justin, B. Lee, Gross, Wilczek, Politzer,
Becchi, Rouet, Stora, Tyutin, Nambu, Goldstone, Higgs, Brout, Englert,
Bouchiat, Iliopoulos, Meyer, and many,
many others. The establishment of nonabelian gauge theory at the
classical and quantum level ranks with the discoveries of special
and general relativity and quantum mechanics as one of the great
achievements of physics.

I have had the privilege of spending the beginning of my
scientific life in Utrecht with Tini Veltman, Gerard 't Hooft,
Bernard de Wit, Hans Reif, and then, after some postdocs, the
rest of my scientific life at Stony Brook with Frank Yang. The
many discussions with them in past and present times have revealed to me
the personal side of their great discoveries, the uncertainties and
worries, but also the satisfaction of just doing interesting
work, and the slow realization that something important was being
constructed. The friendship with them has been and still is a continuing
source of support for my own activities.
Bob Mills I only met at the retirement symposium of Frank in 1999.
He struck me as a very decent and honest person. Unfortunately
he died soon after.

Yang and Mills wrote their pivotal paper in 1954 without any
reference to gravity \cite{YM}. However, already in the 1920's the
ideas of
gauge theory were developed in the context of gravity, notably by
Weyl, and we shall connect these two approaches. We shall show
that one can also apply the gauge field formalism of Yang and Mills
to gravity and \sg, with an action quadratic in curvatures instead
of the linear Einstein-Hilbert action. Of course Weyl in addition
considered a locally scale invariant formulation of gravity in one
of his earlier papers, in order to explain the meaning of
electromagnetism, and this approach (with an extra factor of $i$
later added when quantum mechanics was discovered) led to the
modern concept of gauge symmetry. Also in \sg\ such a theory
exists. It is called conformal \sg, to distinguish it from the
ordinary theories of \sg\ which one might call Poincar\'{e} (or
rather super Poincar\'{e}) theories. Conformal \sg\ has also an
$R^2$ action, very much of the type of Yang and Mills, but one
needs constraints on curvatures and torsions. Ordinary (and also
conformal) \sg\ can be very beautifully written in superspace
\cite{5,GGRS}
but then one also needs constraints on the supercurvatures and
supertorsions, as we shall discuss below.
We begin, however, with a rather elementary introduction intended
for readers who are unfamiliar with \sy\ and \sg.
\section{Basics of \sg}
As with any gauge theory, one can either approach \sg\ by first
studying its coupling to matter, or one can begin by constructing
the gauge action. The gauge field for \sy\ is the spin $3/2$ field
$\j_\m$ which is called the gravitino field. It is clear that this
gauge field should be a vector-spinor field of the form $\j_\m^\a$
($\m=0,3$ and $\a=1,4$) because gauge fields transform as
$\d$(gauge field)$=\pa_\m$(parameter)$+\dots$. For \sy\ the
parameter is a spinor $\e^\a$, so the gauge field should have the
index structure $\j_\m^\a$. It is real (because it is the partner
of the gravitational field, see below), and of course
anticommuting as the spin-statistics relation suggests. As a
vector-spinor, it contains on-shell only helicities $\pm 3/2$, but
off-shell the field $\j_\m^\a$ also contains spin $1/2$ parts,
just like a gauge field $A_\m$ contains on-shell helicities $\pm
1$, and off-shell also helicity zero.

In $3+1$ dimensions, one can have theories with $N=1$ up to $N=8$
real gravitinos, but beyond $N=8$ the massless representations of the
underlying \sy\ algebra contain particles with spin larger than 2,
and no consistent gauge theories exist for these cases.

For the simplest case, $N=1$, the \sy\ algebra has massless
representations in terms of physical states with adjacent
helicities $(J,J+1/2)$. By combining these representations with
representations with helicities $(-J,-J-1/2)$, one obtains the
field content for massless fields with spin $J$ and $J+1/2$. This
is a result of Salam and Strathdee who also pioneered the
superspace approach \cite{3}. The gauge action for $N=1$ \sy\ is based
on
the multiplet with $(J,J+1/2)=(\frac32,2)$. The alternative, spin
$(1,3/2)$, does not lead to a consistent \sg\ theory; it couples
the Maxwell field to gravitinos but the resulting gauge theory
contains no physical particles (the curvatures vanish according to
the field equations of this model). However, one can view the spin
$(1,3/2)$ multiplet as a matter multiplet, and couple it to the
spin $(3/2,2)$ gauge multiplet. The result is the simplest
extended \sg\ theory, the $N=2$ model with spin content
$(1,3/2,3/2,2)$. It realizes Einstein's goal of unifying gravity with
Maxwell
electromagnetism, and has an $O(2)$ symmetry which rotates the two
gravitinos into each other \cite{FN}.
In the same way the $N=8$ model has a local $SO(8)$ symmetry \cite{CJ}.
The group $SO(8)$ is not big enough
to contain the nongravitational symmetry group $SU(3)\times
SU(2)\times U(1)$, and this precluded direct contact with
phenomenology.

Because \sy\ requires that the  gravitino is part of a spin
$(3/2,2)$ multiplet, a gauge theory of local \sy\ necessarily
contains gravity. Thus gauge \sy\ is a theory of gravity, and this
explains the name \sg. One could also arrive at this name by
starting with a matter theory of rigid \sy, and couple it to
gravity. Because the \sy\ parameter $\e^\a$ transforms as a spinor
under local Lorentz transformation, it becomes in general
spacetime-dependent after a local Lorentz transformation, even if it
initially was rigid (spacetime independent), hence also from this
point of view the name \sg\ seems appropriate.

In their pioneering article, Yang and Mills first wrote down
kinematical transformation rules, and only afterwards constructed
a gauge action for the group $SU(2)$. Before them, one usually
began with an action with particular dynamics, and then set out to
describe the symmetries of the dynamical model under
consideration. Likewise in \sg\ one can begin either by first
constructing the gauge action, or the coupling to matter, or both,
and afterwards discuss the symmetries, or one can begin with the
symmetries, and afterwards construct actions with these
symmetries. In \sg\ there is a special problem with symmetries
which does not occur in Yang-Mills theories, the problem of
auxiliary fields. If one studies the transformation laws of the
local symmetries, which are for the $N=1$ \sg\ theory the local
Lorentz and general coordinate symmetry of ordinary gravity and
further local \sy, then one discovers that ``the local gauge
algebra does not close" without auxiliary fields. Namely, the
commutator of two local \sy\ transformations is not only a sum of
local symmetries, but one finds also field equations (of fermionic
fields in general, but not always \cite{T}) 
on the right-hand side~\cite{4}. Thus one does
not have a representation of the local gauge algebra in terms of
fields. In some cases (but not all cases) one can add a few
``auxiliary fields" (fields whose field equations do not describe
physical states) such that the local gauge algebra closes. In the
early literature on \sg, finding a set of auxiliary fields was an
art.

Having a closed gauge algebra allows one to go to superspace.
Superspace is a space with fermionic coordinates $\q^{i\a}$ (with
$i=1,\dots,N$) in addition to the bosonic coordinates. In
superspace one can define supercurvatures and supertorsions, but
contrary to ordinary general relativity, in \sg\ these
supercurvatures and supertorsions must satisfy certain
constraints. These constraints define the geometry; they are
inserted by hand from the outside, and are not field equations.
Again, in the early days finding a correct set of constraints was
an art. (Correct means here: such that no ghosts and no particles
with spin larger than 2 do occur). A translation method was
constructed, called gauge completion, which could map a \sg\
theory which was given in $x$-space, into superspace \cite{AN},
but this was only possible if the local gauge algebra was closed. For this
reason the auxiliary fields, even though not physical, were of
great importance. It should be mentioned that the superspace
approach contains superfields (fields depending on $x^\m$ and
$\q^{i\a}$) which contain many more local symmetries and many more
auxiliary fields then the corresponding theory in $x$-space. The
map from $x$-space into superspace corresponds to a particular
gauge of the full superspace theory. However, working with
non-gaugefixed superfields simplifies the calculations a good deal.

The classical \sg\ theories are gauge field theories. To quantize
them one followed initially the same procedures as followed by 't
Hooft and Veltman and others for Yang-Mills gauge theories.
Namely, one added a gauge-fixing term and corresponding
Faddeev-Popov ghost action. For the gravitinos the most useful
gauge fixing term was ${\cal
L}(\mbox{fix})=\frac{1}{4 \xi}(\bar\j_\m\g^\m)\dr(\g^\n\j_\n)$ with
$\xi=1$ \cite{7}.
Because ghosts have opposite statistics from the
corresponding local gauge parameters, and the parameters for \sy\
are anticommuting (Grassmann variables), the ghosts for local \sy\
were commuting spinors. In addition the ghosts for local Lorentz
invariance $c^{mn}=-c^{nm}$ and the ghosts for general coordinate
transformations $c^\m$, were anticommuting. The ghosts for local
Lorentz symmetry could be eliminated by their algebraic field
equations. However, a new feature in the quantization of gauge
theories was discovered. Certain \sg\ theories contain
antisymmetric tensor fields (for example the \sg\ theories in ten
dimensions), and for these theories, the Faddeev-Popov ghost
actions were themselves gauge actions! Thus one had to do the
Faddeev-Popov procedure all over again, and in this way
ghosts-for-ghosts emerged. To deal with this complicated issue a
general framework was developed by Batalin and Vilkovisky, which
generalized earlier work by Zinn-Justin, and which is nowadays
called the antifield formalism \cite{BV}.
The antifields in this approach
are external fields which satisfy bracket relations with the
original fields; it is a kind of covariant Hamiltonian formalism.
However, the antifields have opposite statistics from the original
field, and consequently the bracket itself is anticommuting.

The antifield formalism made contact with the BRST formalism which
was soon established after 't Hooft and Veltman had renormalized
Yang-Mills theory. The authors (Becchi, Rouet, Stora,
and later Tyutin \cite{BRST}) noticed that after covariant quantization
the
quantum action (the sum of the classical action and the gauge
fixing term and the ghost action) still had a residual rigid
symmetry with a constant anticommuting para\-meter $\Lambda$.
Combining the antifield formalism with BRST symmetry has led to a
profound geometrical framework. It generalizes the work of Dirac,
Heisenberg and Pauli, and Gupta and Bleuler on QED to the nonabelian
level. In earlier studies of $N=1$ \sg\
it had been found that unitarity required four-ghost couplings if
one did  not add auxiliary fields \cite{F}.
The BRST-antifield formalism
shows that four-ghost couplings are a direct consequence of the
nonclosure of the gauge algebra. On the other hand, if one starts
with a classical formulation of \sg\ with auxiliary fields and
with closed gauge algebra, then standard Faddeev-Popov
quantization is applicable (in the simplest models at least), and
eliminating the auxiliary fields from the quantum action, the same
four-ghost couplings are found. This demonstrates the close
connection between gauge algebras, quantization and geometry.

\section{Simple $(N=1)$ supergravity.}
The simplest way to introduce \sg\ is to begin with ordinary
gravity, add the action for a free spin $3/2$ field, couple it to
gravity according to the usual rules of general relativity, and
then try to make the action invariant under local \sy. We shall
follow this procedure, but go back to Herman Weyl's 1929
formulation of gravity \cite{W}, and generalize it to \sg.

According to this approach one begins with the spacetime symmetry
algebra of the tangent frames, the Lorentz algebra with generators
$M_{mn}$ satisfying $[M_{mn},M_{pq}]=\eta_{np}M_{mq}+3$ further
terms. Then one associates a gauge field (connection) to $M_{mn}$,
the spin connection $\o_\m^{\ mn}=-\o_{\m}^{\ nm}$. Finally one
constructs the Yang-Mills curvature for the Lorentz group. In
general a Yang-Mills curvature has the form
\begin{equation}
F_{\m\n}^{\ \ a}=\pa_{\m}\o_\n^{\ a}-\pa_\n \o_\m^{\ a}+ f^a_{\
bc}\o_\m^{\ b}\o_\n ^{\ c}
\end{equation}
and in the case of the Lorentz group one obtains,
using the structure constants of the Lorentz group,
\begin{equation}
R_{\m\n}^{\ \ mn}(\o)=\pa_{\m}\o_\n^{\ mn}-\pa_\n \o_\m^{\
mn}+\o_\m^{\ mp}\eta_{pq}\o_\n^{\ qn}- \o_\m^{\
np}\eta_{pq}\o_\n^{\ qm}
\end{equation}
Gauge transformations read in general
\begin{equation}
\d W_\m^{\
a}=\pa_\m\l^a+gf^{a}_{\ bc} W_\m^{\ b}\l^c
\end{equation}
and become for the Lorentz group
\begin{equation}
\d \o_{\m}^{\ \ mn}=D_{\m}\l^{\ mn}\equiv \pa_\m\l^{mn}+\o_\m^{\
mn}\eta_{pq}\l^{qn}+\o_\m^{\ np}\eta_{pq}\l^{mq}
\end{equation}
The curvatures transform homogeneously,
\begin{equation}
\d(\l)R_{\m\n}^{\ \ mn}=\l^m_{\ p}R_{\m\n}^{\ \ pn}+\l^n_{\
p}R_{\m\n}^{\ \ mp}
\end{equation}

To construct an action Weyl was faced with the problem of
contracting the indices of $R_{\m\n}^{\ \ mn}$. It was natural  to
consider an action linear in $R_{\m\n}^{\ \ mn}$ because that was
Einstein's approach (but an alternative is an $R^2$ action, see
below). To this purpose he considered the vielbein fields $e_\m^{\
m}$ which naturally arise if one tries to put the Dirac action in
curved space. The Dirac matrices satisfy $\{\g^m(x),\g^n(x)
\}=2\eta^{mn}$ in flat space, but in curved space $\{\g^\m,\g^\n
\}=2 g^{\m\n}$ where $g^{\m\n}(x)$ is the metric, and then it is
natural to write $\g^\m(x)=\g^me_m^{\ \m}(x)$ \cite{Wig}.
Latin indices $(m,n)$ correspond to tensors in flat space (the tangent
frames,
the freely falling lifts) while Greek indices $(\m,\n)$ correspond
to coordinates in curved space. Substituting the expression for
$\g^\m(x)$ into the Clifford algebra immediately yields
\begin{equation}
\eta^{mn} e_m^{\ \m} e_n^{\ \n}=g^{\m\n}
\end{equation}
Thus Weyl constructed the Einstein-Hilbert action for gravity in
terms of the spin connection and vielbein fields
\begin{equation}
{\cal L}=-\frac12e R_{\m\n}^{\ \ mn}(\o)e_m^{\ \n}e_n^{\ \m}
\end{equation}
where $e=\det e_\m^{\ m}$, and $e_m^{\ \n}$ is the matrix inverse
of $e_{\n}^{\ m}$. Straightforward algebra shows that it is equal
to the action Einstein and Hilbert had written down in terms of
metrics $g_{\m\n}$ and Christoffel connections $\Gamma_{\m\n}^{\ \
\r}(x)$ \cite{AE}. However, for the couplings to fermions, one needs
Weyl's formulation.

For the extension of Weyl's approach to \sg\ it is useful to
consider the Poincar\'{e} algebra instead of the Lorentz algebra with
$[M_{mn},P_l]=\eta_{nl}P_m-\eta_{ml}P_n$ and $[P_m,P_n]=0$. We
associate again $\o_{\m}^{\ mn}$ with $M_{mn}$, but $e_\m^{\ m}$
can now be associated with $P_m$. For our purposes we must
generalize this to an anti-de Sitter algebra, where instead of
$[P_m,P_n]=0$ one has $[P_m,P_n]=\a^2 M_{mn}$ with $\a$ a free
parameter. The Yang-Mills curvatures now become
\begin{align}
R_{\m\n}^{\ \ mn}(\o,e)&=R_{\m\n}^{\ \ mn}(\o) +
\a^2(e_\m^{\ m}e_\n^{\ n}-e_\m^{\ n}e_{\n}^{\ m})\\
R_{\m\n}^{\ \ m}(\o,e)&=\pa_\m e_\n{}^{m}-\pa_\n
e_\m{}^{m}+\o_\m^{\ mp}\eta_{pq}e_\n^{\ q}-\o_\n^{\
mp}\eta_{pq}e_\m^{\ q}
\end{align}

In order to generalize the Poincar\'{e} algebra to a superalgebra
which can be used for \sg, one needs anticommuting generators. The
spin-statistics connection suggests that these parameters should
be spinors. Spinors can be described in a four-component formalism
or in a two-component formalism\footnote{Soon after a tensor
calculus was established for general relativity, Ehrenfest in
Leiden sent a letter to van der Waerden in G\"{o}ttingen, asking
if something similar could be done for spinors. These result is
``the van der Waerden formalism" of two-component dotted and undotted
spinor indices $A,\dot{A}$ \cite{BvdW}.}.
Although two-component spinors are widely used,
and form the irreducible representations of the Lorentz
group, we shall first use four-component spinors to reach a
wider audience. We consider generators $Q^\a$ ($\a=1,..\,,4$). They
transform as spinors under the Lorentz group,
$[M_{mn},Q^\a]=\frac14\ll[\g_m,\g_n] \rr^\a_{\ \b}Q^\b$, and they
are constant in space and time, $[Q^\a,P_m]=0$. If \sy\ is to map
bosons into fermions, and fermions into bosons, there should be no
kernel for $Q^{\alpha}$ (the null space of $Q^{\alpha}$ should be trivial). 
Hence $\{Q^\a,Q^\b \}$ should be equal to an
operator which has no kernel. Covariance and the fact that the
only commuting generators available are $P_m$ and $M_{mn}$, allows
then only
\begin{equation}
\{Q^\a,Q^\b \} = \g^{m,\a\b}P_m+\a'
\ll\g^{[m}\g^{n]}\rr^{\a\b}M_{mn}
\end{equation}
In fact, for the superPoincar\'{e} algebra $\a'=0$, but for the
super-anti de Sitter algebra $\a'$ is equal to the parameter $\a$
we already encountered. Consistency (satisfying the Jacobi
identities) then also requires $[Q^\a,P_m]=\a(\g_m)^\a_{\
\b}Q^\b$. The corresponding Yang-Mills curvatures are
\begin{align}
R(M)= R_{\m\n}^{\ \ mn}(\o,e,\j)&=R_{\m\n}^{\ \
mn}(\o)-\alpha \bar{\j}_\m\g^{[m}\g^{n]}\j_{\n}+
\alpha^2(e_\m^{\ m}e_\n^{\ n}-e_\m^{\ n}e_\n^{\ m})\\
R(P)=R_{\m\n}^{\ \ m}(\o,e,\j)&=R_{\m\n}^{\ \
m}(\o,e)+\bar{\j}_\m\g^{m}\j_{\n} \label{n12}\\
R(Q)=R_{\m\n}^{\ \ \a}(\o,e,\j)&=\ll D_\m(\o)\j_\n+\frac12\a e_\m^{\
m}\g_m\j_\n\rr-\m\rightarrow\n
\label{n13}
\end{align}
The gauging of superalgebras has been discussed in \cite{9}. 
\footnote{To
check signs and the Jacobi identities, an easy method is to assume that
the curvature two-forms vanish, and then to check that the exterior
derivative of them also vanishes. Note that the two forms
$\bar \psi  \psi $ and $\bar \psi  \gamma^{mnp}\psi $ vanish for 
Majorana
spinors.}

Minkowski space can be viewed as a coset space, namely Poincar\'{e}
algebra/Lorentz algebra, and anti-de Sitter space is the coset
space $SO(3,2)/SO(3,1)$. Superspace is in the same way the
coset space $\{M_{mn},P_m,Q^\a\}/\{M_{mn}\}$, namely
super-Poicar\'{e} algebra/Lorentz algebra, or super-anti de Sitter
algebra/Lorentz algebra. The curvatures of the coset generators
are usually called torsions. According to this terminology,
the Lorentz curvature is a
genuine curvature, but $R_{\m\n}^{\ \ m}$ and $R_{\m\n}^{\ \ \a}$
are torsions.

Having come so far, it is natural to follow Yang and Mills and
construct an action for \sg,  quadratic in curvatures, and
invariant under the two local spacetime symmetries (local \sy\
will be discussed later). One can still contract the indices of
the curvatures in various ways with vielbein
fields. But there is one way which uses constant tensors, just as
Yang and Mills used in their paper, and that is by using
$\e$-tensors \cite{8}
\begin{equation}
{\cal L}= R_{\m\n}^{\ \ mn}R_{\r\s}^{\ \
pq}\e^{\m\n\r\s}\e_{mnpq}+a(\bar{R}_{\m\n})_\a(\g_5)^\a_{\
\b}(R_{\r\s})^\b\e^{\m\n\r\s}
\label{n14}
\end{equation}
where $(\bar{R}_{\m\n})_\a = (R_{\m\n}{}^\beta)^\dagger i
(\gamma^0)^\beta{}_\alpha$
with $(\gamma^0)^2=-1$, and $(\gamma_5)^2=+1$, while
$a$ is a constant to be fixed later. Note that
\begin{itemize}
\item[(i)] parity is preserved
\item[(ii)] the action is a density (because the tensor
$\e^{\m\n\r\s}$ with entries $\pm 1,0$ is a density)
\item[(iii)] no term with the square of $R_{\m\n}^{\ \ m}$ can be
constructed in this way.
\item[(iv)]  all fields have ``geometrical dimensions'', meaning
that the gravitational coupling constant has been absorbed into
the gravitino.
So there will be no gravitational coupling constants in any of the
formulas below, but note that $\alpha$ and $a$ are dimensionful.
The gravitino has dimension 1/2 ,and $\alpha$ and $a$ have
dimension 1.
\end{itemize}

Now comes a surprise: substituting the expression for $R_{\m\n}^{\
\ mn}(\o,e,\j)$ into ${\cal L}$, the leading term
\begin{equation}
\e^{\m\n\r\s}\e_{mnpq}R_{\m\n}^{\ \ mn}(\o)R_{\r\s}^{\ \ pq}(\o)
\end{equation}
is a total derivative \cite{8}. (In form language it reads
$R^{mn}(\o)\wedge R^{pq}(\o)\e_{mnpq}$.  Under a variation
$\o\rightarrow \o+\d\o$ with arbitrary $\d\o$, one finds $\d
R^{mn}=D\d\o^{mn}$ and then partial integration yields a vanishing
result due to the Bianchi identity, $DR=0$. Equivalently, locally $d
(R\wedge R)=D R\wedge R+ R\wedge DR=0$, hence $R\wedge R$ is locally a
total derivative. These formulas hold for any spin connection
$\omega$, whether it is an independent field or a dependent field).
The cross terms in the bosonic sector yield the
Einstein-Hilbert action\footnote{Use that
$\e_{mnpq}\e^{\m\n\r\s}=e_m^{\ \m}e_n^{\ \n}e_p^{\ \r}e_q^{\
\s}+23$ other terms, due to antisymmetrization in $\m,\n,\r,\s$.}
but in the formulation of Weyl
\begin{equation}
R_{\m\n}^{\ \ mn}(\o) e_\r^{\ p}e_\s^{\
q}\e_{mnpq}\e^{\m\n\r\s}\sim eR(e,\o)
\end{equation}
while the four vielbein fields yields a cosmological constant
\begin{equation}
(e_\m^{\ m}e_\n^{\ n}e_\r^{\ p}e_\s^{\
q})\e^{\m\n\r\s}\e_{mnpq}\sim  e
\end{equation}

In the fermionic sector one now finds similar results \cite{8}:
\begin{itemize}
\item the leading term $\ll D_\m\bar{\j}_\n\g_5D_\r\j_\s \rr
\epsilon^{\m\n\r\s}$
cancels the cross term $\alpha R(M)\, \bar \psi \gamma_{mn}\psi $
in $R(M) R(M)$ up to a total derivative.
\item the cross terms in $R(Q)R(Q)$
yield the gauge action for the gravitinos
\begin{equation}
{\cal L}_{3/2}=(\bar{\j}_\m\g_\n)\g_5 (D_\r\j_\s
)\e^{\m\n\r\s}\sim\bar{\j}_\m\g^{\m\r\s}D_\r\j_\s
\end{equation}
(by $\g^{\m\r\s}$ we mean $\g^\m\g^\r\g^\s$ antisymmetrized in
$\m,\r,\s$). This is the action Rarita and Schwinger first wrote down
(up to a field redefinition as in (35)) 
when they studied nuclear beta decay under the assumption that
neutrinos have
spin 3/2  \cite{RS}.
\item the remaining term is the masslike term
$(\bar\j_\m\g_\n\g_5\g_\r\j_\s)\e^{\m\n\r\s}\sim
\bar\j_\m\g^{\m\n}\j_\n$, which is needed in the presence of a 
cosmological
constant in order that local supersymmetry is preserved so
that the gravitino remains massless \cite{DZF}.
\end{itemize}

The action is manifestly invariant under local Lorentz and
Einstein transformations (general coordinate transformations).
What can now be said about the local \sy\ of this action? The
Yang-Mills transformation rules for local \sy\ follow directly
from the super anti-de Sitter algebra. Applying the general rules
of Yang and Mills but with the structure constants of the super anti-de
Sitter algebra yields\footnote{Again a simple way to check these
results is to write the transformation rules as one-forms and
to verify that the curvature two-forms transform
into each other.}
\begin{align}
&\d e_\m^{\ m} = -\bar\e\g^m\j_\m \label{18}\\
&\d\j_\m=\pa_\m\e-\frac{\a}{2} e_\m^{\ m}\g_m\e+\frac14\o_\m^{\
mn}\g_{mn}\e \label{19}\\
&\d\o_\m^{\ mn}= \a\bar\e \g^{mn}\j_\m
\label{20}
\end{align}
The action is {\bf not} invariant under these transformation
rules. The reason is, of course, that the curvatures in (\ref{n14})
were contracted with constants which are not invariant tensors
under local supersymmetry.
At this point a new subtlety arises which is
absent in Yang-Mills theories: constraints are needed on the
curvatures to obtain the correct law for $\d \o_\m^{\ mn}$.

Variation of the action under local \sy\ transformations (an easy
task since curvatures rotate into curvatures\footnote{The
transformation rules of the curvatures are obtained by replacing
the connections by their corresponding curvatures in (19-21).})
leads to $R(Q)R(M)$ terms which cancel if
$a=8\a$. However,
a term proportional to $R(P)R(Q)$ is left. Since $R(P)=0$ in
(\ref{n12}) can be
algebraically solved (see below) but $R(Q)=0$ in (\ref{n13}) can not be
algebraically solved, we impose  the torsion constraint
\begin{equation}
R_{\m\n}^{\ \ m}(P)=0.
\label{21}
\end{equation}
Its solution yields $\o_{\m}^{\ \ mn}$ as a function of $e_\m^{\
m}$ and $\j_\m$,
\begin{equation}
\o_\m^{\ mn}=\o_\m^{mn}(e)-\frac12
(\bar\j_\m\g^m\j^n+\bar\j^m\g_\m\j^n-\bar\j_\mu\g^n\j^m)
\end{equation}
where $\o_{\m}^{\ mn}(e)$ is the spin connection in terms of
vielbein fields which one can find in text books on general
relativity. The terms with $\bar\j\g\j$ are torsion. Torsion was
introduced into general relativity in the 1920's by E. Cartan
\cite{Car}, but it has found its natural realization in supergravity.
The constraint in (\ref{21})
is also the field equation of the spin connection in $N=1$ $x$-space
supergravity, see (26),
but in superspace or conformal \sg\ the
constraints are not field equations.
One can now determine $\d\o_\m^{\ mn}(e,\j)$ by applying the chain
rule and using~(\ref{18}) and~(\ref{19}). Then one can check the
invariance of the action by varying all fields. This is laborious
(and the way it was first done); in particular, the variation of
$\o(e,\j)$ leads to a lot of terms if one uses the chain rule. The
crucial observation, arrived at much later, is that the $\o$ field
equation (whose solution is $\o=\o(e,\j)$ with a complicated
$\o(e,\j)$) is identically satisfied, once one uses everywhere
$\o(e,\j)$ instead of $\o$. The reason is that after substituting
$\o(e,\j)$ for $\o$, the variation of the composite object
$\o(e,\j)$ is always multiplied by $\d S/\d\o$ which is zero.
Hence, one can forget about the variation of $\o(e,\j)$ altogether,
provided one works in second-order formalism with a dependent
field $\o$.

It is amusing to see how elegantly this all works out. From
$R(P)=0$ one finds that the extra variation
$\Delta \omega^m{}_n$ is determined by
\begin{equation}
\delta R(P)^m= -\bar \epsilon \gamma^m R(Q)
+ \Delta \omega^m{}_n e^n =0
\end{equation}
> From this expression one can solve for $\Delta \omega^m{}_n$.
The total variation of the spin connection is then
$\delta \omega^m{}_n + \Delta \omega^m{}_n $,
and this expression agrees with the result of applying the chain
rule to (23) and using (19) and (20).

It is also easy to see that the sum of all variations
due to $\delta \omega^m{}_n + \Delta \omega^m{}_n $
cancels. Variation of the action, and partial integration
of the Einstein-Hilbert term yields the following result
in form notation
\begin{equation}
\delta {\cal L} =  2 (\delta \omega^{mn} + \Delta \omega^{mn})
[ \epsilon_{mnpq} DR(M)^{pq}
+ a \bar R(Q)\gamma_5 \frac14 \gamma_{mn} \psi ]
\end{equation}
Using $DR(\omega)=0$ and $R(P)=0$, the terms inside the square
bracket cancel provided again $a=8\alpha$.

Originally, Freedman, Ferrara and the author began with $\o(e)$,
and found successive terms in the action and transformation laws
by computer until they obtained an invariant action \cite{1}.
Two thousand terms had to cancel, and did cancel.
Looking back,
it is now clear that these extra terms just replaced $\o(e)$ by
$\o=\o(e,\j)$. Clearly then, imposing the constraint $R(P)=0$,
which replaces $\o$ by $\o(e,\j)$ at the beginning, is an enormous
simplification.

One can also work with an independent field $\o$. This is called
first-order formalism. Then one has no constraint $R(P)=0$, and
one must find $\d(\e)\o$ by direct means. This can be done, and
was done by Deser and Zumino [2]. Also this result is complicated, and
not equal to the transformation laws found in the second order
formalism.

In fact, one can even more clearly show where the choice between
first-order and second-order formalism is made. The variation of
the action with respect to $\o$ can be written as
\begin{equation}
\d S \sim\int \e^{\m\n\r\s}\e_{mnrs}R_{\m\n}^{\ \ m}(P)[\hat{\d}\o_\r^{\
nr}-\Omega_\r^{\ nr}(e,\j)]e_\s^{\ s}
\end{equation}
where $\Omega_\r^{\ nr}(e,\j)$ is a complicated function
and $\hat\d\o_\r^{\ nr}$ denotes any variation $\o_\r^{\ nr}$.
We only used
$\d e_\m^{\ m}=-\bar{\epsilon}\g^m\j_\m $ and $\d\j_\m=D_\m(\o)\e$ to 
obtain this
result. The coefficient of $\hat\d\o$ is, of course, the $\o$
field equation, according to the general Euler-Lagrange
variational principle. The fact that $\d S$ factorizes is
nontrivial and leads to the choice between first- and second-order
formalism. Second-order formalism puts $R_{\m\n}{}^m(P)=0$, thus
replacing the independent field $\o_\m^{\ mn}$ by the dependent
field $\o_\m^{\ mn}(e,\j)$. This is the result of \cite{1}.
First-order formalism puts $\d\o$ equal to $\Omega$ and this
yields the result of \cite{2}.

In an early article, Volkov and Soroka \cite{6}
gauged the super Poincar\'{e} algebra (not the super
anti de Sitter algebra), but did
not prove its invariance under \sy. They used a first-order
formalism and found $\d\o=0$. This agrees with (\ref{20})
in the limit $\alpha=0$.
As we have explained, this result is incorrect.
They implicitly
assumed that their action would be invariant, and concluded that
supergravities exist for any $N$. Careful study, using $\d\o\neq
0$, shows that $N\leq 8$.

The fact that in second-order formalism one need not take into
account the variations of $\o=\o(e,\j)$ is sometimes called
1.5-order formalism because it combines in some sense the virtues
of first- and second-order formalisms. Namely, one keeps the
composite object $\o=\o(e,\j)$ as one object (not expanding it in
terms of $e$ and $\j$), and this is like first-order formalism.
But then one uses that $\d\o(e,\j)$ is multiplied by $\d S/\d\o$
which is identically zero, and this is due to using second-order
formalism. So, in the end only
$\d e_\mu^m=-\bar\e\g^m\j_\mu$ and $\d\j_\mu=D_\mu\e$ are
needed.
This makes the proof of the supersymmetry invariance of the action for
$N=1$ supergravity very easy, as easy as the gauge invariance of the
action of Yang and Mills.

\section{Covariant quantization.}
Next we briefly discuss the covariant quantization of \sg. We
consider a simplified case, with only external gravitational
fields and no Einstein action. To preserve covariance at the
quantum level, one may use a background field formalism. The spin
$3/2$ action has then a torsionless spin connection
\begin{align}
&{\cal L}_{3/2} = -\frac12 e\bar\j_\m \g^{\m\r\s}D_\r\j_\s\\
&D_\r\j_\s=\pa_\r\j_\s+\frac14\o_{\r mn}(e)\g^m\g^n\j_\s
\end{align}
It is invariant by itself under local \sy\ transformations,
without adding the Einstein action, if the gravitational
background fields are Ricci flat ($R_{\m\n}=0$). We use a \sy\
gauge fixing term which preserves the classical spacetime
symmetries
\begin{equation}
{\cal L}(\mbox{fix})=\frac14 e\bar\j_\m\g^\m \dr \g^\n\j_\n
\end{equation}
The Faddeev-Popov action for the \sy\ ghosts is
\begin{equation}
{\cal L}(\mbox{ghost})=-e\bar{b}_\a\dr c^\a
\end{equation}
where $b^\a$ and $c^\a$ are 4-component commuting Majorana spinors
(or, more precisely, $c^\a$ is a Majorana spinor but $b^\a$ is $i$ 
times a
Majorana spinor). To obtain ${\cal L}(\mbox{fix})$ in the
action, one starts from a gauge fixing term $\d[\g^\m\j_\mu-F]$ in the
path integral, and then one inserts unity into the path integral
as follows
\begin{equation}
I=\int dF e^{\bar{F}\dr F}(\det \dr)^{-1/2}
\end{equation}
Integration over $F$ brings ${\cal L}(\mbox{fix})$ in the action,
but exponentiating $(\det \dr)^{-1/2}$ with Nielsen-Kallosh ghosts
\cite{NK} yields
\begin{equation}
{\cal L}(NK)=-\frac{e}{2}\bar{A}\dr A -\frac{e}{2}\bar{B}\dr C
\end{equation}
where $A$ is a Majorana anticommuting ghost and $B$ and $C$ are
Majorana commuting ghosts. The Faddeev-Popov ghosts remove the
unphysical longitudinal and timelike parts of $\j_\m$, which
correspond to the gauge symmetry $\d\j_\m=\pa_\m\e$ as in QED. The
Nielsen-Kallosh ghosts, on the other hand, remove the spin $1/2$ parts
$\g\cdot\j$ from the spectrum.

The sum of the classical spin $3/2$ action and the gauge fixing
term leads to
\begin{equation}
{\cal L}_{3/2}=\frac{e}{4} \bar\j_\m\g^\s\dr\g^\m\j_\s
\end{equation}
This is a complicated action. One can, however, reduce it to the
Dirac action by some simple field redefinitions. We do this in $n$
dimensions. We use in $n$ dimensions as gauge fixing term
\begin{equation}
{\cal L}(\mbox{fix})=\frac{n-2}{8}e\bar\j_\mu\g^\m\dr\g^\n\j_\n
\end{equation}
This term vanishes for $n=2$, as does the classical action in (27).
We then choose a new basis for the spin $3/2$ field
\begin{equation}
\c_\m=\j_\m-\frac12\g_\m\g\cdot\j,\quad
\j_\m=\c_\m-\frac{1}{n-2}\g_\m\g\cdot\c
\end{equation}
One finds that the action in $n$ dimensions on the basis $\c_\m$
becomes a sum of Dirac actions
\begin{equation}
{\cal L}_{3/2}+{\cal
L}(\mbox{fix})=-\frac{e}{2}\bar\c_\m\dr\c^\m=-\frac12
e\bar\c_m\dr\c^m
\end{equation}
where $\chi_{m} = e_{m}{}^{\mu} \chi_{\mu}$. 
The operator $\dr$ contains now both a term with the spin
connection which acts on the spinor index, and another term
which acts on the vector index of $\c^m$.
This form of the spin 3/2 action has been used to compute chiral
anomalies using quantum mechanics \cite{AGW}.
\section{Conformal simple supergravity.}
We now apply the formalism to a second, far less trivial, example:
conformal simple  \sg\ \cite{10}.
Simple means again that there is one
ordinary \sy\ generator $Q_\a$ ($\a=1,4$), but there is also a
conformal-\sy\ generator $S_\a$. (There also exist extended conformal supergravities, but only with $1\leq N \leq 4$, unlike ordinary supergravities 
for which $1\leq N \leq 8$.) 
The bosonic conformal algebra
contains the translation generators $P_m$ ($m=0,3$) which now
commute with each other and with $Q_\a$, but there are also
conformal boost generators $K_m$, Lorentz generators $M_{mn}$ and
scale generators $D$. The pair $P_m,Q_\a$ resembles the pair
$K_m,S_\a$; for example, also$[K_m,K_n]=0$ and $[K_m,S_\a]=0$. One
might expect that the total set of generators consist of $
\{P_m,M_{mn},D,K_m \}$ and $\{Q_\a,S_\a \}$. However, in the
superalgebra one needs one more bosonic generator $A$ for chiral
transformations of $Q_\a$ and $S_a$. In the literature the
corresponding symmetry is called $R$ symmetry.

Conformal symmetry is believed to be the symmetry of fundamental
interactions at ultra-high energies where masses can be neglected.
It is a larger symmetry than only scale invariance, and whether at
some deep level Nature has conformal invariance is an open
question. According to the criteria of perturbative quantum
field theory, conformal gravity is not viable
as a low-energy effective action because
its propagator has double poles and violates unitarity \cite{L}.
However,  gravity theories are inherently nonperturbative, so maybe 
this is
not the whole story.
In string theories conformal, or rather superconformal,
symmetry plays  a big role, but we shall here not go further into
the physics of conformal symmetry. Rather, we want to construct a
gauge action, hoping that one day it will be used.

All generators have a $Z$ grading: if two generators have grades
$p$ and $q$, their commutator contains only generators with grade
$p+q$. The generators together with their grades are as follows:

\begin{equation}
\left\{
\begin{array}{ccccc}
&&D,A&&\\ K_m & S_\a& \ M_{mn}& Q_\a &P_m\\
-2& -1& 0& 1& 2. \end{array} \right.
\end{equation}
This explains why $[Q_\a,P_n]=0$ and $[P_m,P_n]=0$ (idem for
$K_m,S_\a$). In addition, $A$ commutes with the bosonic conformal
algebra. We shall not write the superconformal algebra down in
detail. It corresponds to $SU(2,2|1)$, which may be defined by
$5\times 5$ matrices with in the $4\times 4$
left-upper part the bosonic $SU(2,2)\sim
SO(4,2)$ conformal algebra, and further the generator $A$ is
represented by a diagonal matrix (with entries $(1,1,1,1,+4)$ up
to an overall scale). In the fifth row and fifth column one finds
$Q_\a$ and $S_\a$. All matrices are supertraceless, $\mbox{str}
M=\sum\limits_{i=1}^4M^i_{\ i}-M^5_{\ 5}=0$.

To gauge this algebra, we introduce again for each generator a gauge
field and a local gauge parameter. These gauge fields and
parameters we denote by
\begin{equation}
f_\m^{\ m}(\xi_K^{\ m}),\v_\m^{\ \a} (\e_S^{\ \a}),b_\m (\l_D),
A_\m (\l_A),\o_{\m}^{\ mn}(\l^{mn}),\j_\m^{\ \a}(\e_Q^\a),e_\m^{\
m}(\xi_P^{\ m}).
\end{equation}
The corresponding curvatures are denoted by
\begin{equation}
R_{\m\n}^{\ \ m}(K),\ \ R_{\m\n}^{\ \ \a}(S),\ \  R_{\m\n}(D),\ \
R_{\m\n}(A),\ \ R_{\m\n}^{\ \ mn}(M),\ \ R_{\m\n}^{\ \ \a}(Q),\ \
R_{\m\n}^{\ \ m}(P).
\end{equation}
We must now construct an action quadratic in curvatures which
preserves parity. In the vein of Yang and Mills we allow only
constants to contract indices, but no vielbein fields. The action
should be invariant under {\bf all} local symmetries except
$P$-gauge transformations, and under Einstein transformations. We
expect again that we shall need constraints on the curvatures to
achieve this, and we shall in a methodical way deduce these
constraints, and solve them. In superspace one also finds
constraints on curvatures and torsions.

Invariance under local scale transformations allows only products
with curvatures with opposite grade, since the grades are
proportional to the scale ($D$ eigenvalue). Hence, we only
consider the products $R(K)R(P), R(S)R(Q)$, and bilinears in
$R(M),R(D),R(A)$. Since the chiral weights of $R(Q)$ and $R(S)$
are opposite, $([A,Q_\a]=+c(\g_5)_\a^{\ \b}Q_\b$ while
$[A,S_\a]=-c(\g_5)_\a^{\ \b}S_\b$ with $c$ a constant, local
chiral invariance is then also achieved. Local Lorentz invariance
requires to contract the curvatures with Lorentz-invariant tensors
such as $\e^{\m\n\r\s}$, $(\g_m)^\a_{\ \b}$, $(\g_5)^\a_{\ \b}$.
This will get us $D,A,M$ invariance of the action. Parity
restricts the coupling of $R(A)$ to only $R(D)$ (because $A$ has
negative parity, and $D,M$ have positive parity). The most general
action then reads
\begin{equation}
S=\int d^4x\e^{\m\n\r\s}[R_{\m\n}^{\ \ mn}(M)\e_{mnrs}R_{\r\s}^{\
\ rs}(M)+\a R(Q)^\a_{\m\n}(\g_5)_{\a\b}R(S)^\b_{\r\s}+
\b R_{\m\n}(A)R_{\r\s}(D)]
\label{38}
\end{equation}
with $\a,\b$ constants. No $R(P)R(K)$ coupling is possible which
preserves parity. This is analogous to the observation that no term
quadratic in $R(P)$ was allowed in ordinary supergravity.

Before going on, we mention that the only curvatures which can
lead to constraints which can be algebraically solved are
$R(P),R(Q),R(M)$ and $R(D)$. They contain terms with products of a
vielbein $e_\m^{\ m}$ times another gauge field. Since we can
invert the vielbein, we can eliminate the other gauge field.
So ahead of time we know that our formalism should only lead to
constraints on these curvatures.

We now study whether the action in (\ref{38})
is invariant under the symmetries of the superconformal algebra
using as transformation laws those of Yang and Mills,
but with the structure constants of the superconformal algebra.
We recall that the variation of the action only involves curvatures,
as curvatures rotate into curvatures. We begin with the symmetries
with negative grade. Invariance under local $K$ and $S$
transformations leads to two constraints on the curvatures
and fixes the free parameters in the action
\begin{equation}
R_{\m\n}^{\ \ m}(P)=0,\ \ \ \ R_{\m\n}^{\ \ \a}(Q)\sim
\e^{\m\n\r\s}R_{\r\s}(Q)^\b(\g_5)_\b^{\ \a}, \ \ \ \a=8,\ \ \
\b=-4i.
\end{equation}
This is indeed the complete action, but we need further
constraints. In a conformally invariant action (in fact, scale
invariance is enough for this argument) the fields with positive
scale weight cannot appear in the kinetic part of the action, as
there are no dimensionful constants which can make the action
scale invariant. This explains ahead of time that $f_\m^{\ m}$,
$\v_\m^{\ \a}$ must be eliminated by constraints. It comes perhaps
not as a surprise that also $\o_{\m}^{\ mn}$ will be eliminated
just as in ordinary supergravity. Hence, at the end the fields
$f_\m^{\ m}$ (conformal vielbein), $\v_\m^{\ \a}$ (conformal
gravitino) and $\o_{\m}^{\ mn}$ (spin connection) will completely
have been eliminated, leaving only $e_\m^{\ m}$ (graviton),
$\j_{\m}^{\ \a}$ (gravitino) and $A_\m$ (chiral gauge field) as
physical fields. A special role will be played by $b_\m$;
conformal symmetry (with local parameter $\xi_K^{\ m}$) acts on
$b_\m$ like $\d b_\m=\xi_k^{\ m}\eta_{mn}e_\m^{\ n}$, and hence
$b_\m$ can be gauged away by a suitable $K$-gauge choice.
Equivalently: in the action $b_\m$ cancels.

So now we more or less know what we should find, and we shall now
derive these results in a deductive manner, by studying the
algebra (kinematics) rather than the action (dynamics).
The first problem to face is that if there are constraints
and one solves them by expressing one field in terms of others,
this dependent field no longer transforms according to the
(structure constants of the) superalgebra,
but rather according to a result which follows from the chain rule.
We already saw this in the case of the spin connection in simple 
ordinary supergravity.
All constraints are invariant under all
local symmetry transformations as given by the superalgebra, except
under local $P$ and local $Q$ transformations.
Thus, even after imposing and solving the constraints all dependent
fields still transform as before imposing the constraints
under all local symmetries except  $P$ and $Q$.
This proves that the action is invariant under local
$K$, $S$, $M$, $D$, $A$ transformations.
We shall replace the requirement
that the action be invariant under local $P$ transformations
by the requirement that it be invariant under general coordinate
(Einstein) transformations. Requiring invariance both under
$P$ and under  Einstein transformations would be too much;
however there is a deep relation between both as we now explain.

There exists a systematic procedure to find the constraints
themselves. The basic idea is the observation that a general
coordinate transformation differs from a gauge transformation by
a curvature term
\begin{eqnarray}
\delta(\xi^\n) \omega_\mu^a
&\equiv&
\partial_\mu \xi^\nu \omega_\nu^a +
\xi^\nu \partial_\nu \omega_\mu^a  \cr
&=& \partial_\mu(\xi\cdot \omega^a)
+ f^a{}_{bc}\, \omega_\mu^b \xi\cdot \omega^c
-\xi^\nu (\partial_\mu \omega_\nu^a - \partial_\nu \omega_\mu^a
  + f^a{}_{bc}\, \omega_\mu^b \omega_\nu^c)\cr
  &=& D_\mu(\xi\cdot \omega^a) -\xi^\nu R_{\mu\nu}^a
\label{neq}
\end{eqnarray}
We require the theory to be Einstein invariant, as well
as gauge invariant
under all symmetries {\it except} $P$. (We do not require invariance
both under Einstein and under $P$-gauge symmetries
in order not to overcount.)
If we consider now the commutator of two gauge transformations other 
than
$P$, and produce a $P$-gauge transformation on the right hand side, we 
still
have  a symmetry if the difference of this $P$ gauge transformation
and an Einstein transformation vanishes. This difference is a curvature
according to (\ref{neq}), and in this way one deduces the constraints
on curvatures.

The only
commutators which produce $P$ generators on the right-hand side,
but which do not involve $P$ on the left-hand side, are (see the
grade table) only the $\{Q,Q \}$ anticommutator. So we shall
compute $\{Q,Q \}\omega_\m{}^a$ taking for $\omega_\m{}^a$ first the
gauge field with highest grade $(e_\m^{\ m})$, then the next
$(\j_\m^{\ \a})$ etc.

We begin with
\begin{equation}
[\d(\e_1^{\ Q}),\d(\e_{2}^{\ Q})]e_\m^{\ m}=\d_P\ll
\frac12\bar\e_2^{\ Q}\g^m\e_1^{\ Q}\rr e_\m^{\ m}.
\end{equation}
Since $\d(\e^Q)e_\m^{\ m}\sim\bar\e\g^m\j_\m$ and $e_\m^{\ m}$ and
$\j_\m^{\ \a}$ are physical fields, there will be no corrections:
$\Delta(\e_Q)e_\m^{\ m}=0$ and $\Delta(\e_Q)\j_\m^{\ \a}=0$. On
the other hand, we need a term $(1/2)\bar\e_2^{\ Q}\g^\l\e_1^{\
Q}R_{\l\m}^{\ \ m}(P)$ on the right-hand side to convert $\d_P$
into an Einstein transformation. This leads to the first
constraint
\begin{equation}
R_{\m\n}^{\ \ m}(P)=0.
\end{equation}
It can be solved to give $\o_{\m}^{\ mn}=\o_{\m}^{\ \
mn}(e,\j,b)$. (The field $b_\m$ cancels in the action, but not in
$\o_\m^{\ mn}.$) This analysis is the same as in the previous
section. In particular we find in the variation of $\o(e,\j,b)$ a
correction term $\Delta \o$.
Since $R(P)$ rotates into $R(Q)$, the correction term
$\Delta \omega$  is proportional to $R(Q)$.

Next we consider the \sy\ commutator on $\j_\m^{\ \a}$. Since
there is an extra $\Delta\o_\m^{\ mn}$, one gets
\begin{equation}
[\d(\e_1^{\ Q}),\d(\e_{2}^{\ Q})]\j_\m^{\ \a}=\d_P\ll
\frac12\bar\e_2^{\ Q}\g^m\e_1^{\ Q}\rr \j_\m^{\ \a}-\left\{ \ll
\frac14\g_{mn}\epsilon_2^{\ Q}\rr\Delta(\epsilon_1^{\ Q})\o_\m^{\ mn}
- \epsilon_1^{\ Q} \leftrightarrow \epsilon_2^{\ Q}\right\}
\end{equation}
We need on the right-hand side $(1/2)\bar\e_2^{\ Q}\g^\l\e_1^{\
Q}R_{\l\m}(Q)$ according to (42), but we have got the expression 
$-\{\dots\}$. Now
$\Delta\o_\m^{\ mn}$ is proportional to $R(Q)$ but it does not
give the exact $R(Q)$ term we need. The difference is a new
constraint
\begin{equation}\label{29}
\g^\m R_{\m\n}(Q)=0 \qquad\qquad (\g^\m\equiv \g^m e_m^{\ \m}).
\end{equation}
It can be solved, and then eliminates the conformal gravitino,
$\v_\m^{\ \a}=\v_\m^{\ \a}(e,\j,A,b)$. Hence there in now also a
correction $\Delta\v_\m$ in the transformation law of the
conformal gravitino. The constraint in~(\ref{29}) is an
irreducibility constraint; it removes a part from $R_{\m\n}(Q)$
with lower spin. It implies also the second constraint we found before,
and can also be written as an antisymmetry condition
\begin{equation}
R_{\m\n}(Q)+\frac12 \e_{\m\n}^{\ \ \r\s}\g_5 R_{\r\s}(Q)=0,
\qquad\qquad \g_{[\m}R_{\n\r]}(Q)=0 .
\end{equation}

Next we consider the \sy\ commutator on $b_\m $ or $A_\m$. Since
$\d(\e^Q)b_\m=(1/2)\bar\e\v_\m$ and $\d
A_\m=-i\bar\e^Q\g^5\v_\m$, we find in these commutators an extra
$\Delta\v_\m^{\ \a}$ because we already eliminated $\v_\m$. We can
then compute what $\Delta \v_\m^{\ \a}$ should be in order that
one gets an Einstein transformation. This yields
\begin{equation}
\Delta(\e^Q)\v_\n=\frac12\g^\m\e^QR_{\m\n}^{\ \
D}+\frac{i}{4}(\g_5\g^\m\e)R_{\m\n}(A).
\end{equation}
On the other hand, we could directly compute
$\d_{gauge}(\e^Q)\v_\m(e,\j,A_\m)$ by the chain rule, and, subtracting
$\d_{\mbox{gauge}}(\e^Q)\v_\m$, we would then find what
$\Delta(\e^Q)\v_\m$ really is. The difference between the
$\Delta\v_\m$ which we need and the actual result for
$\Delta\v_\m$ is then the new constraint. The direct evaluation
of $\Delta\v_\m$ is laborious,and a much simpler method is to vary
instead the constraint $\g^\m R_{\m\n}(Q)=0$, and to find
$\Delta(\e^Q)\v_\m$ by requiring that this variation vanishes. We
did this already for $\Delta\o_\m^{\ mn}$ where we varied $R(P)$.
Variation of $\g^\m R_{\m\n}(Q)=0$ leads then to
\begin{align}\label{412}
\Delta\v_\n =\frac16\g^\m\e R_{\m\n}(D)+\frac{1}{24}\g_\n^{\
\m\l}R_{\m\l}(D)-\frac14
(\g^\m\g_5\e)R_{\m\n}(A)-\frac{i}{16}(\g_\n\g^{\
\m\l}\g_5\e)R_{\m\l}(A)\nonumber\\
-\frac{1}{12}\g^\m\g_{mn}\e R_{\m\n}^{\ \
mn}(M)-\frac{1}{48}\g_\n^{\ \m\l}\g_{mn}R_{\m\lambda}^{\ \ mn}(M).
\end{align}

Laborious but straightforward algebra then yields for
$\Delta\v_\m$ (needed) $-\Delta\v_\m$ (as in~\ref{412}) the
following new constraint
\begin{equation}\label{33}
R_{\m\n}^{\ \ mn}(M)e_n^{\ \n}e_{m\r}+R_{\m\r}(D)+\bar\j^\l\g_\m
R_{\r\l}(Q) =0.
\end{equation}
The term with
$R(Q)$ term is a ``supercovariantization"; it is present
to remove $\pa_\m\e$ terms in the  variation of this constraint.
(Since $\Delta\o\sim\e\g R(Q)$ and $R_{\m\n}^{\ \ mn}(M)\sim
\pa_\m \o_\n-\pa_\n\o_\m$, the $R(Q)$ terms in~(\ref{33}) have the
structure $\bar\j\g R(Q)$.) By taking the part antisymmetric in
$\m\r$ one finds a duality constraint
\begin{equation}
R_{\m\n}(D)+\frac{i}{4}\e_{\m\n}^{\ \ \r\s}R_{\m\n}(A)=0
\end{equation}
(we used $\g_{[\m}R_{\n\r]}(Q)=0$).

Since  there are no independent fields left, there are no further
constraints. Since all constraints are invariant under
$K,S,M,D,A$ and Einstein symmetry, the transformation laws for
these symmetries are without correction terms $\Delta h_\m^{\ A}$.

We conclude that the action of conformal $N=1$ \sg\ is invariant
under $K,S,M,D,A,Q$ and Einstein gauge transformations.
The action is given in (\ref{38}). The
physical fields are $e_\m^{\ m}, \j_\m^{\ \a}, A_\m$ and $b_\m$,
but $b_\m$ cancels in the  action. The constraints are
\begin{equation}\label{415}
\left\{
\begin{array}{l}
R_{\m\n}^{\ \ m}(P)=0,\quad \g^\m R_{\m\n}(Q)=0,\\
R_{\m\n}(M)+R_{\m\n}(D)+\bar\j^\l\g_\m R_{\n\lambda}(Q)=0,\\
R_{\m\n}(D)+\frac{i}{4}\tilde{R}_{\m\n}(A)=0.
\end{array}\right.
\end{equation}
These constraints are field equations in ordinary \sg,
but here they are only constraints, not field
equations. They are imposed by hand from the outside and define
the theory. So it makes a difference wether one imposes them or not, and
we do impose them.
The ``Einstein equation" implies the duality
constraint $R(D)\sim\tilde{R}(A)$, and this enables one to write
the Yang-Mills action for $A_\m$ in an affine form as
$\e^{\m\n\r\s}R_{\m\n}(A)R_{\r\s}(D)$.
In one respect simple conformal supergravity is almost too simple:
one does not need auxiliary fields.
\section{Constraints in superspace}
The geometry of Einstein space is given by vielbein fields
$e_\m^{\ m}$ and connections $\o_{\m}^{\ \ mn}$. In \sg\ one
uses supervielbein fields
\begin{equation}
E_{\Lambda}^{\ M}(\x,\q);\ \ \ \ \Lambda=(\m,\a),\ \ \ \ M=(m,a)
\end{equation}
so there are both curved fermionic indices $\a=1,4$ and flat
fermionic indices $a=1,4$. The connection is a supervector, but it
is only Lorentz-algebra valued
\begin{equation}
\Omega_{\Lambda}^{\ mn}=(\Omega_{\m}^{\ mn},\Omega_{\a}^{\ mn})
\end{equation}
In terms of these two geometrical objects one can define
supertorsions $T_{MN}{}^{P}$ and supercurvatures $R_{MN}^{\ \ mn}$
as follows\footnote{In early work, Arnowitt, Nath and Zumino
\cite{ANZ}
considered a kind of super Lorentz algebra, with generators
$M^{MN}$ instead of $M^{mn}$. They wrote down Einstein equations
where all vector indices became supervector indices, but this
theory contained higher spin fields and ghosts, and was abandoned.
In \sg\ one only allows the Lorentz group.} \cite{5,GGRS}
\begin{equation}
\{{\cal D}_M,{\cal D}_N\}=T_{MN}{}^{P}{\cal D}_P+R_{MN}^{\ \
mn}J_{mn}
\end{equation}
where ${\cal D}_M=D_M+\Omega_M^{\ mn}J_{mn}$, with $J_{mn}$ the
Lorentz generators, and $D_M=(\pa_m,D_\a)$. The derivative $D_\a$
is the rigid supersymmetry derivative
$D_\a=\frac{\pa}{\pa\q^\a}-\q^\b\g^\m_{\b\a}\frac{\pa}{\pa x^\m}$.
We switch at this point from 4-component flat spinor indices
$a=1,4$ to 2-component flat spinor indices $A=1,2$ and
$\dot{A}=1,2$. The following is a list of constraints which yield
$N=1$ \sg. First there are constraints which express the
connections into the supervielbeins, and also the bosonic
supervielbein $E_m^{\ \Lambda}$ in terms of the fermionic
supervielbein $E_\a^{\ \Lambda}$
\begin{align}
& T_{AB}^{\ \ C}=0,\quad T_{A\dot{B}}^{\ \ \ \dot{C}}=0\\
& T_{mn}^{\ \ r}=0,\quad T_{A\dot{B}}^{\ \ \ m}=-i\s^m_{A\dot{B}}\\
&T_{Amn}(\bar\s^{mn})^{\dot B}_{\ \dot C}=0
\end{align}
(the matrices $\bar\s^{mn}$ are Lorentz generators).
Then there are so-called representation-preserving constraints
\begin{equation}
T_{AB}^{\ \ \dot{C}}=T_{AB}^{\ \ m}=0
\end{equation}
They are needed to be able to define ``chiral superfields",
superfields which contain matter (quarks, Higgs bosons, leptons).
They are defined by ${\cal D}_A\bar\f=0$, and so one has also
$\{ {\cal D}_B,{\cal D}_A \}\bar\f=0$,
and since $J^{mn}\bar\f=0$ on a scalar superfield,
one needs for consistency the representation preserving
constraints. Finally there are the conformal (sometimes called
nonconformal) constraints.
They turn the superspace theory from a conformal supergravity
into a superPoincar\`e or super anti-de Sitter  supergravity.
There are various ways of choosing
these. One way is
\begin{equation}
T_{Am}^{\ \ n}=0.
\end{equation}

The transformation laws of tensors in curved space or superspace
are usually treated with tensor calculus, according to which a
field at $x$ is related to a field at $x'$, for example
\begin{equation}
g_{\m\n}'(x')=\frac{\pa x^\r}{\pa x'^\m}\frac{\pa x^\s}{\pa
x'^\n}g_{\r\s}(x)
\end{equation}
However, there exists a formalism which is much more like Yang-Mills
theory, due to Siegel and Gates \cite{GGRS}.
In Yang-Mills theory a gauge transformation of a scalar
field is given by
\begin{equation}
\f'(x)=e^{U(x)}\f(x),\qquad U(x)=U^a(x)T_a
\end{equation}
where $T_a$ are constant matrices acting on the indices of
$\f(x)$, and $U^a(x)$ are the arbitrary local gauge parameters.
For \sg\ one can write a general supercoordinate transformation as
$Z\rightarrow Z'=Z'(Z)$ where $Z=(x^\m,\q^\a)$. A scalar field
transforms then as
\begin{align}
&\f'(Z)=e^{U(Z)}\f(Z) \label{57}\\
&U(Z)=U^\a(Z)D_\a+U^m(Z)\  \pa/\pa x^m
\end{align}

The usual definition of the transformation rule
of a scalar field is
\begin{equation}
\f'(Z')=\f(Z)
\end{equation}
but if one writes $\f(Z)$ as $\f(Z)=e^{-U(Z)}\f'(Z)$ by
inverting~(\ref{57}) one finds a more familiar expression
\begin{equation}
\f'(Z')=e^{-U(Z)}\f'(Z)
\end{equation}
In particular
\begin{equation}
Z'=e^{-U(Z)}Z\equiv Z'(Z)
\end{equation}
This formula gives the relation between the Yang-Mills parameters
$U(Z)$ for superEinstein transformations, and the more conventional
parametrization
$Z'=Z'(Z)$.

In this way one can write superdiffeomorphisms (or ordinary
diffeomorphisms) as Yang-Mills transformations; the only
difference with internal symmetries is that one has replaced the
matrices $T_a$ of Yang and Mills by operators $\frac{\pa}{\pa
x^\m}$ and $D_\a$. Using this formalism, one can solve the
constraints on the supertorsions, and find the unconstrained
prepotentials which form the starting point for a path integral
description of the quantum theory \cite{GGRS}.

\end{document}